\documentclass[a4paper,12pt]{article}
\usepackage{graphicx}
\usepackage{color}
\usepackage{cite}
\usepackage[auth-sc-lg]{authblk}
\usepackage{fullpage}
\begin{document}

\title{Ellipsometry with polarisation analysis  at cryogenic temperatures inside a vacuum chamber}
\author{S. Bauer$^{1,a}$, B. Grees$^1$, D.~Spitzer$^{1}$, M.~Beck$^{1,c}$, R.~Bottesch$^1$, H.-W.~Ortjohann$^1$, B.~Ostrick$^{1,2}$, T.~Sch\"afer$^1$, H. H. Telle$^3$,
A.~Wegmann$^{1,b}$, M. Zbo\v{r}il$^1$ and C.~Weinheimer$^1$}

\affil[1]{ Institut f\"ur Kernphysik, Westf\"alische Wilhelms-Universit\"at M\"unster,
D-48149 M\"unster, Germany}
\affil[2]{Institut f\"ur Physik, Johannes-Gutenberg-Universit\"at Mainz, D-55128 Mainz, Germany}
\affil[3]{Department of Physics, College of Science, Swansea 
University, Singleton Park, Swansea SA2 8DU, United Kingdom}
\affil[a]{corresponding author s.bauer@uni-muenster.de}
\affil[b]{present address: Max-Planck-Institut f\"ur Kernphysik, D-69117 Heidelberg, Germany}
\affil[c]{present address: Institut f\"ur Physik, Johannes-Gutenberg-Universit\"at Mainz, D-55128 Mainz, Germany}
\maketitle
\begin{abstract}
\textit{
In this paper we describe a new variant of null ellipsometry to determine thicknesses and optical properties of thin films on a substrate at cryogenic temperatures. In the PCSA arrangement of ellipsometry the \underline{p}olarizer and the \underline{c}ompensator are placed before the \underline{s}ubstrate and the \underline{a}nalyzer after it. Usually, the polarizer and the analyzer are rotated to find the intensity minimum searched for in null ellipsometry. In our variant  we rotate the polarizer and the compensator instead, both being placed in the incoming beam before the substrate. Therefore the polarization analysis of the reflected beam can be realized by an analyzer at fixed orientation.
We developed this method for investigations of thin cryogenic films inside a vacuum chamber, where the analyzer and detector had to be placed inside the cold shield at a temperature of $T \approx 90$~K close  to the substrate. All other optical components were installed at the incoming beam line outside the vacuum chamber, including all components which need to be rotated during the measurements.
Our null ellipsometry variant  has been tested with condensed krypton films on a highly oriented pyrolytic graphite substrate (HOPG) at a temperature of $T \approx 25$~K.  
We show that it is possible to determine the indices of refraction of condensed krypton and of the HOPG substrate as well as thickness of krypton films with reasonable accuracy.}
\end{abstract}
%
%
%

%
%
\section{Introduction}
\label{sec:intro}
Since Alexandre Rothen described the "Ellipsometer" in 1945~\cite{Rot45}, the technique has been developed into a well known procedure to measure film thicknesses and to determine optical properties of a film. Ellipsometry applies light of a well-defined state of polarization that is reflected from the investigated multilayer system. After reflection the state of polarization is analyzed. 
One application of the technique is the determination of the refractive index of a gas condensed on a surface at low temperature~\cite{Kruger}. In this case, the film must not contain impurities and the layer system has to be enclosed in a vacuum chamber.

In the past many different variations of ellipsometry set-ups were realized \cite{Azz87}. Because of the arrangement of \underline{p}olarizer, \underline{c}ompensator, optical \underline{s}ystem and \underline{a}nalyzer, this commonly used constellation is called "PCSA" arrangement (see figure~\ref{fig:pcsa}). In standard applications so-called null-ellipsometry is used: Here the polarizer and the compensator produce elliptically  polarized light such that after the reflection at the multilayer system the light is fully linearly polarized. Thus it can be extinguished by an analyzer.

\begin{figure}[!!h]
	\centering
\includegraphics[width=0.6\textwidth]{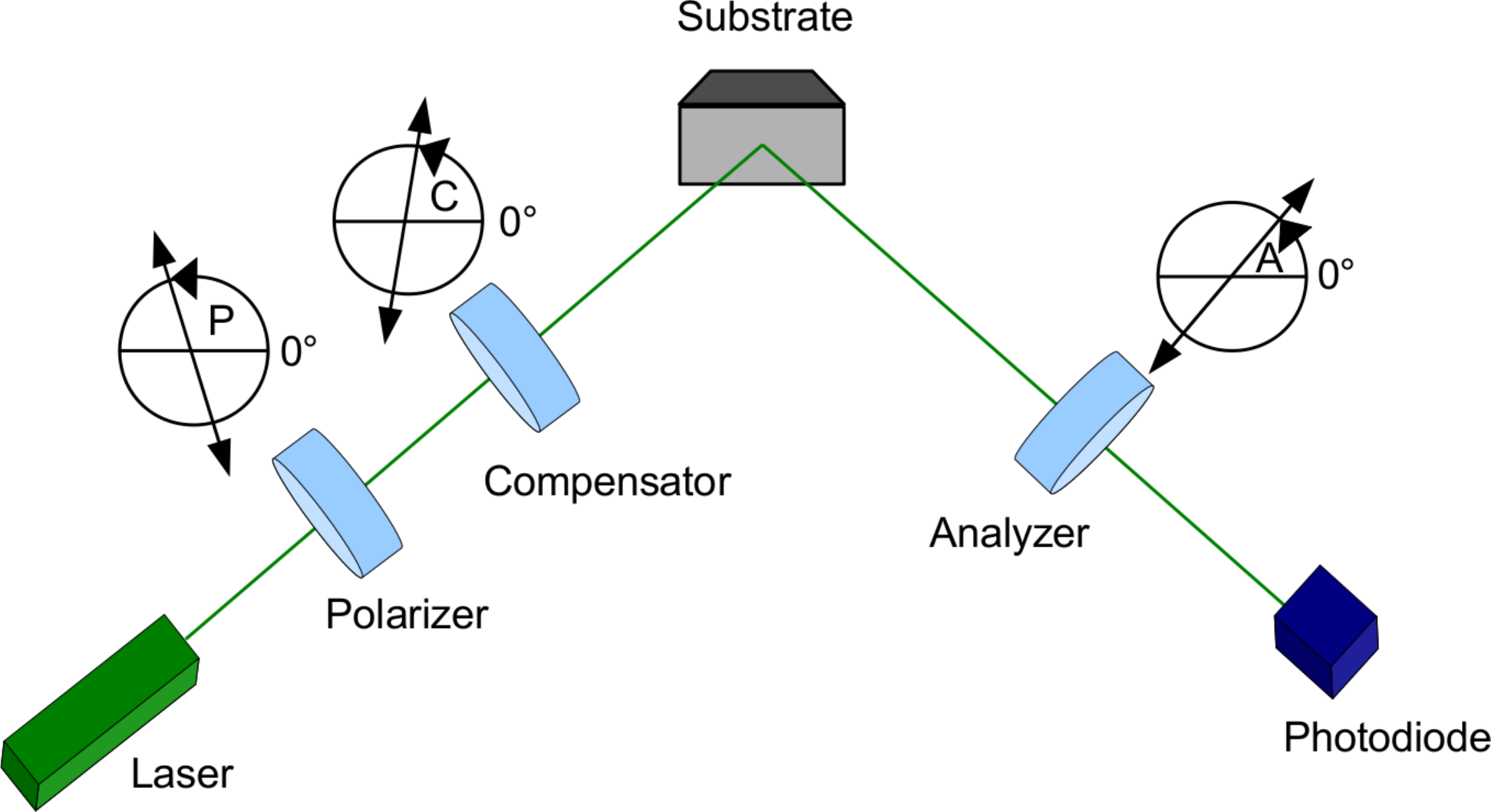} 
  \caption{Ellipsometry set-up in PCSA arrangement with polarizer and compensator in the incoming beam line and with analyzer and light detector in the beam line reflected from a substrate. The angles of the polarizer, the fast axis of the compensator, and the analyzer with respect to the plane of incidence are named $P$, $C$ and $A$ in the following.}
    \label{fig:pcsa}
\end{figure}

\begin{figure}[!!h]
	\centering
  \includegraphics[scale=0.4]{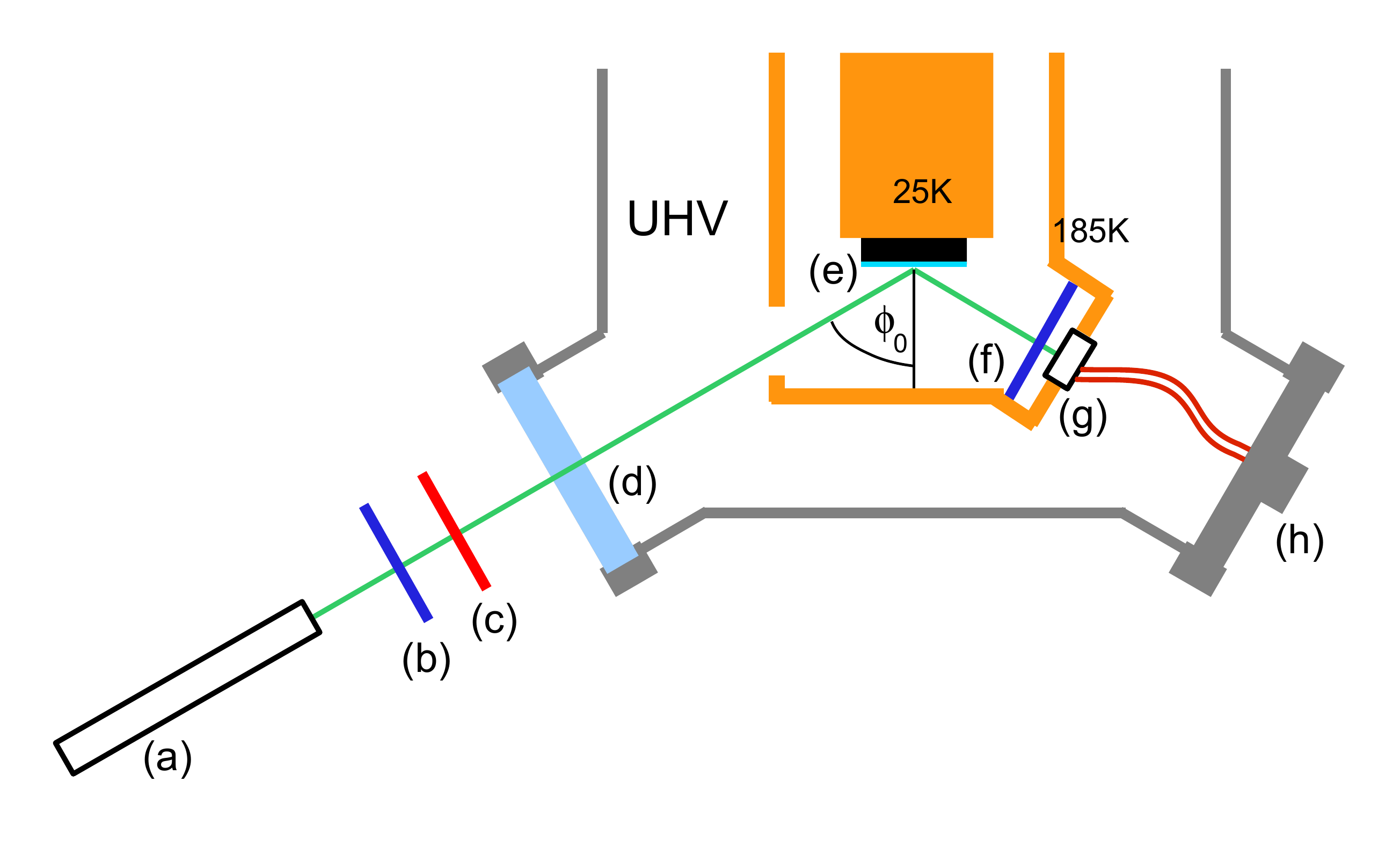} 
  \caption{Ellipsometry set-up in PCSA arrangement with analyzer and detector in an ultra-high vacuum chamber. The set-up comprises a (a) light source, (b) polarizer, (c) compensator, (d) vacuum window, (e) multilayer system, (f) analyzer, (g) photodiode detector and (h) electric feedthrough with current amplifier. In our set-up the analyzer and the detector are placed inside the ultra-high vacuum chamber at cryogenic
temperatures. For clarity, the inner cold shield at 12~K surrounding the substrate is not shown.}
  \label{fig:1}
\end{figure}

In order to measure thin films with sub-monolayer resolution in a cryogenic ultra-high vacuum environment, a high resolution null-ellipsometry variant has been explored in this work. When measuring thin cryogenic films, using standard experimental arrangements all optical elements remain outside the vacuum chamber at room temperature before and after the multilayer system (e.g. \cite{McM04,Kraus05}). Whenever this is not possible the method presented here is an interesting alternative. Such a situation could arise if the reflected light after passing the multilayer system would have to be guided out of the vacuum chamber over a large distance before entering the analyzer section. The reason for this could be -- like in our case case -- a not perfectly flat substrate causing a too large divergence of the reflected beam. To circumvent this problem, the reflected light needs to be analyzed and detected inside the vacuum chamber. If then the analyzer cannot be rotated because of limited space or too low temperatures, the analyzer has to be fixed at a certain angle and the ellipsometry has to be applied in a modified manner.

Our particular PCSA implementation addresses these technical constraints by rotating the compensator in addition to the polarizer and searching for the intensity minimum with a fixed analyzer orientation. As far as we know an ellipsometry variant with fixed analyzer and rotating compensator was presented to be possible in \cite{Azz87} but we haven't found applications or results of such an ellipsometer in literature\footnote{In addition to this variant, a second method solely using a rotating polarizer has been explored~\cite{Knorr89}. Another variant of ellipsometry is the rotating-compensator Fourier ellipsometer described by Hauge et al.; they utilise a rotated compensator in combination with fixed polarizer and analyzer. \cite{Hauge75}.} . 

It will be shown that this method allows the determination of arbitrary film thickness like the standard PCSA method with rotating analyzer and polarizer. Specifically, we apply this variant of standard PCSA ellipsometry at our conversion electron calibration source for the KATRIN neutrino mass experiment \cite{KDR04}:
We condense the krypton isotope $^{83m}$Kr on a highly-oriented pyrolytic graphite (HOPG) substrate at cryogenic temperatures ($25~K$) under ultra-high vacuum conditions~\footnote{Although we usually condense for this application film thicknesses of less than a monolayer, we use ellipsometry with \AA -resolution to monitor the cleanliness of the substrate after laser ablation over typical measurement periods of several days. A stable and clean surface is needed to guarantee a conversion electron energy stability and reproducibility of a few 10~meV .}. This electron source is positioned inside a superconducting split-coil magnet in a LN$_2$ cooled ultra-high vacuum environment. The reflected light would have to be guided out of the vacuum chamber over a distance of about 2~m. Unfortunately, this is not possible due to the beam divergence
caused by the polycrystalline structure of the HOPG substrate\footnote{In a predecessor neutrino mass experiment at Mainz similar PCSA ellipsometry has been applied to determine thicknesses of deuterium and tritium films \cite{bornschein2003, aseev2000}. But because of the problems with the divergent out-going beam the film thickness could only be determined in an offline position before and after the typically two weeks long measurements inside a superconducting magnet.}. Therefore  the
light has to be analyzed and detected inside the vacuum chamber at cryogenic temperatures.

%
%
%
\section{Theoretical description}
\label{sec:theorie}

\begin{figure}
	\centering
	\includegraphics[scale=0.65]{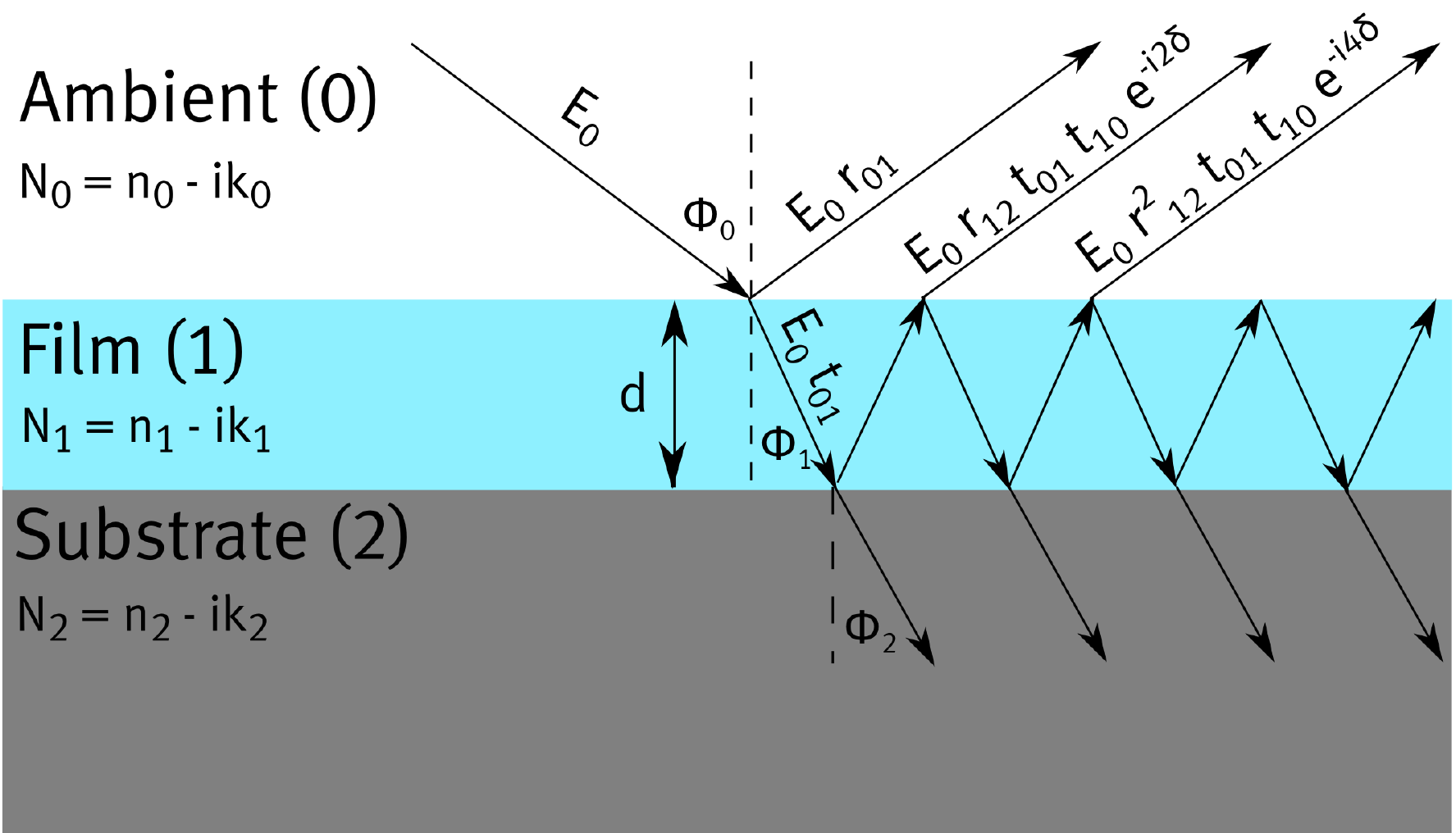}
	\caption{Multiple beam interference at a double layer system consisting of a substrate (index of refraction $N_2$) and a film (index of refraction $N_1)$ with thickness (d) inside an environment (index of refraction $N_0$, typically
  vacuum with $N_0 = 1$). The light beam has an  angle of incidence ($\Phi_0$) with respect to the normal.}
	\label{fig:pcsa}
\end{figure}

In the following we consider a double layer system consisting of a flat substrate, covered by a homogeneously thick  film of thickness $d$ (see fig. \ref{fig:pcsa}). Following the notation of \cite{Azz87, Hea91} we describe a dielectric medium by a complex refractive index $N = n - ik$. With the definition of a plane wave in z-direction of $E(t,z)=E_0 \cdot e^{i(\omega t - kz)}$ $n$ becomes the index of refraction and $k$ the extinction coefficient. Thus the absorption coefficient $\alpha$ can be expressed by the extinction coefficient ($k$) and the vacuum wavelength ($\lambda_0$) as $\alpha=\frac{4\pi k}{\lambda_0}$. 

We assume that the film can be described by an refractive index $N_1 = n_1$ with only a negligible extinction $k_1 \approx 0$.
Also for the ambient (usually vacuum or a gas atmosphere) we assume a real index of refraction $N_0 = n_0$. Only the substrate is described by a complex index of refraction $N_2 = n_2 - ik_2$.
The reflection properties of this system are given by the complex reflection coefficients $R_s$ and $R_p$ for the multilayer system \cite{Azz87}
\begin{equation}
R_s=\frac{r_{01s}+r_{12s}e^{-2i\delta}}{1+r_{01s}r_{12s}e^{-2i\delta}}~~~~~\mbox{and}~~~~~R_p=\frac{r_{01p}+r_{12p}e^{-2i\delta}}{1+r_{01p}r_{12p}e^{-2i\delta}}, 
\end{equation}
with $r_{01s}, r_{12s}, r_{01p}, r_{12p}$ being the coefficients of single reflection for the various interfaces. These are calculated using the Fresnels formulas for the interfaces between the ambient medium and the film (index "01") or between the film and the substrate (index "12") for perpendicular (denoted "s") and parallel (denoted "p") polarized waves with respect to the plane of incidence. The coefficients $r_{01s}, r_{12s}, r_{01p}, r_{12p}$ depend on the indices of refraction of the regarded interface and the angle of incidence $\Phi_0$. For a vacuum wavelength $\lambda$ of the laser, the film thickness $d$ causes a phase shift $\delta$:

\begin {equation}    
	\delta=\frac{2\pi d}{\lambda}\sqrt{N_{1}^2-N_{0}^2\sin^2{\Phi_0}} 
	        = \frac{2\pi d}{\lambda}\sqrt{n_{1}^2-n_{0}^2\sin^2{\Phi_0}} \, .
\end{equation}
As usual $P$ and $A$ constitute the rotation angles of the polarizer and analyzer as defined by the orientation of transmitted polarized light with respect to the plane of incidence (see figure \ref{fig:pcsa}). $C$ is the angle between the fast axis of the compensator and the plane of incidence. If the compensator is a quarter wave plate
the light intensity behind the analyzer is given by  \cite{Azz87}
\begin{eqnarray}
  I  \propto |&R_p \cos(A) [\cos(C) \cos(P-C) +i \sin(C) \sin(P-C)] \\ \nonumber
                &+ R_s  \sin(A) [\sin(C) \cos(P-C) -i \cos(C) \sin(P-C)] |^2 \, .
	 \label{eqn:intensitaet}
\end{eqnarray}
Two of the three angles $P$, $C$, $A$ are defined by the condition of null intensity $I = 0$, 
while the third one can be chosen freely in most cases. 
Then the properties of the system are described by   
\begin{equation}
	\rho_S = \frac{R_p}{R_s} = -\tan(A) \frac{\tan(C) -i \tan(P-C)}{1 +i \tan(C) \tan(P-C)}.
	\label{eqn:Grundgln}
\end{equation} 

Usually the compensator angle is set to $C = \pm \pi/4$ resulting in
\begin{equation}
\rho_S = \mp \tan(A) e^{\mp 2i(P \mp \frac{\pi}{4})}~~~~\mbox{for}~~~~C=\pm \frac{\pi}{4}.
	\label{eqn:pa_elli}
\end{equation}

In the following we call this method "PA ellipsometry", in which the compensator is fixed to $C=\pm \pi/4$ while the the polarizer and the analyzer are varied in order to find an intensity minimum.

\begin{figure}[!!!b]
	\centering
	\includegraphics[scale = 0.44]{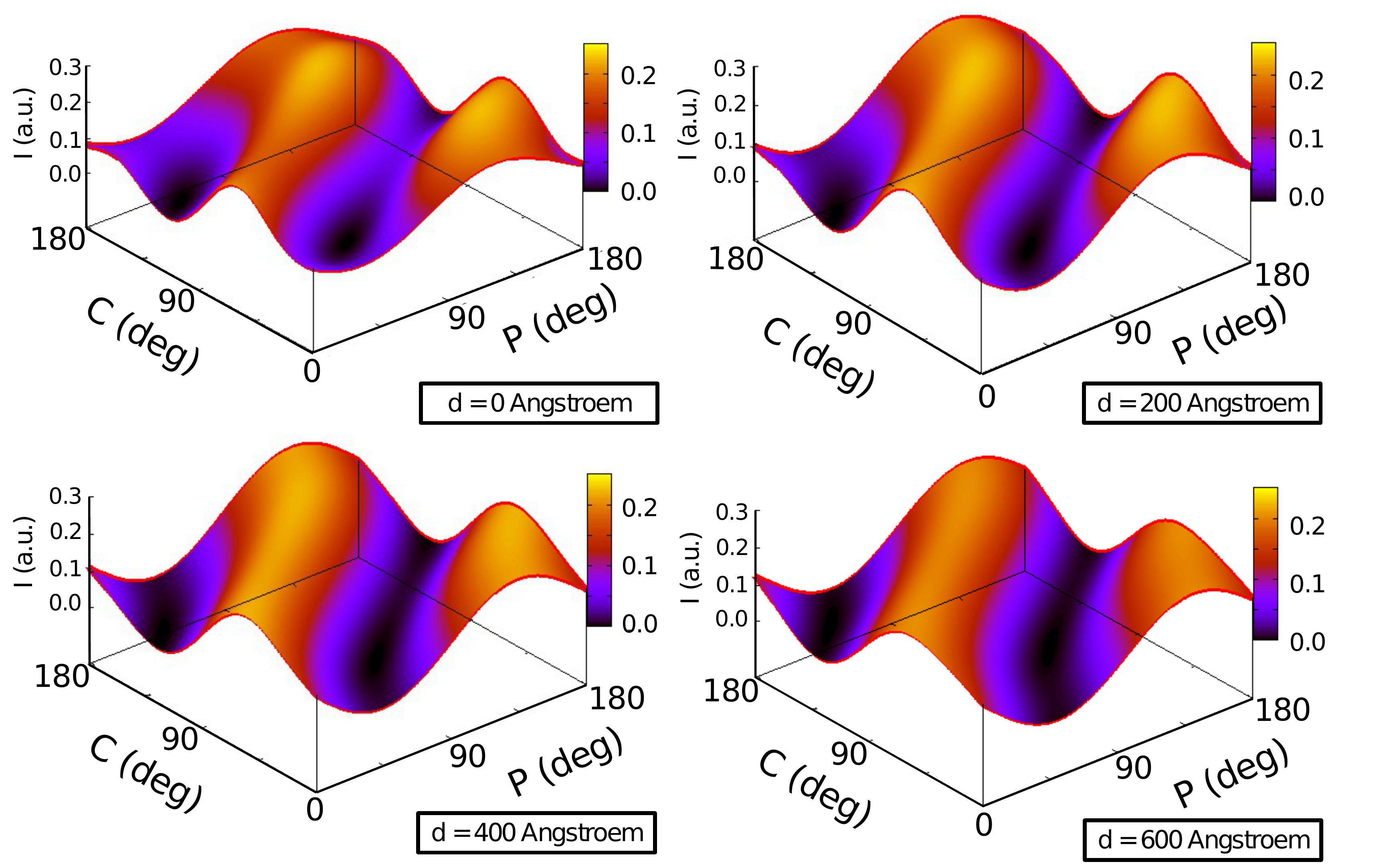}
	\caption{Intensity profile for a full scan of polarizer and compensator angles $P$ and $C$ for various
	film thickness  $d = 0$~\AA~(upper left), $d = 200$~\AA~(upper right), $d = 400$~\AA~(lower left) and 
	$d = 600$~\AA~(lower right). The analyzer angle was set to $A=30^\circ$ with respect to the plane of incidence. The simulations were made assuming a krypton film on a HOPG substrate in a vacuum 
	environment ($N_0=1$, $N_1=1.38$, $n_2 = 2.61$, $k_2=1.55$). The laser wavelength in these simulations was set to $\lambda = 543.5$~nm (green HeNe laser) and the angle of incidence to $\Phi_0 = 60^\circ$.}
	\label{fig:pc_simulations} 
\end{figure}

For more detailed information we refer to reference~\cite{Azz87}. 
We would like to note that in the literature the complex variable $\rho_S$ is often expressed as
\begin{equation}
  \rho_S = \tan({\Psi})Â e^{i \Delta}.
\end{equation}
where $\tan({\Psi})$ corresponds to the absolute value and $\Delta$ to the complex phase of $\frac{R_p}{R_s}$ in the polar expression. In the modified variant of PCSA ellipsometry, which was outlined in the introduction, the analyzer remains fixed at a certain angle $A$ and the compensator angle $C$ is varied. We still consider the compensator to be a quarter-wave plate. Therefore, equation (\ref{eqn:Grundgln}) still holds but not equation (\ref{eqn:pa_elli}).
In order to prove that we can obtain null intensity by varying the polarizer and compensator 
angles $P$ and $C$ we plot the intensity as function of $P$ and $C$ for different krypton film thicknesses $d$
in figure~\ref{fig:pc_simulations}.

It shows indeed, that for all film thicknesses regions with null-light intensity can be found. But figure 
\ref{fig:pc_simulations} also illustrates that the minima might be flat, thus limiting the sensitivity.
Like in PA ellipsometry we see again two minima, now in the $PC$-plane\footnote{ Of course,  figure \ref{fig:pc_simulations} only shows some exemplary simulations to demonstrate these statements, but it is not a full mathematical proof.}. 
We are determining the minimum of the light intensity as function of the polarizer and compensator angles $P$ and $C$ for a fixed analyzer angle $A$. From now on we call this method "PC ellipsometry". 
Via equation (\ref{eqn:Grundgln}) the angles $P$ and $C$ define a complex variable $\rho_S$.
In order to facilitate the determination of the corresponding film thickness $d(P,C)$ we define
two corresponding angles $\tilde A$ and $\tilde P$.

Similar to Euler`s representation, we generally can express any complex number $\rho_S$ by two angles $\tilde A, \tilde P \in [0, \pi [$ as
\begin{equation}
	\rho_S = \tan(\tilde A) \cdot e^{i 2 (\tilde P+\pi/4)} \, .
	\label{eqn:complex}
\end{equation}  
Therefore we can translate our angles $P$ and $C$ defining the intensity minimum for a given film thickness $d$
into the corresponding angles $\tilde A$ and $\tilde P$ via:
\begin{equation}
	\tan(\tilde A) \cdot e^{i 2 (\tilde P+\pi/4)} = \rho_S 
	           = -\tan(A) \frac{\tan(C) -i \tan(P-C)}{1 +i \tan(C) \tan(P-C)} \, .
	\label{eqn:pseudo_trans}
\end{equation} 
The left-hand side of equation (\ref{eqn:pseudo_trans}) looks identical to equation (\ref{eqn:pa_elli})
for $C=-\pi/4$.  This means, if a film thickness $d$ would be characterized by a pair ($P$,$C$) in PC ellipsometry, the transformation with equation~(\ref{eqn:pseudo_trans}) describes which angles $\tilde P$ and $\tilde A$ would have been encountered in standard PA ellipsometry for the same film thickness. The advantage of this transformation is that one can use the same data analysis tools as for PA ellipsometry.

\section{Experimental set-up}
\label{sec:set_up}
A schematic overview of the experimental set-up is shown in figure~\ref{fig:1}~\footnote{The real set-up uses two cold shields - an inner cold shield at 12~K and an outer cold shield at 90~K. For a better overview the inner cold shield was neglected in figure~\ref{fig:1}.}. The light source is a HeNe laser ($\lambda = 543.5$~nm, P$_{laser} = 0.5$~mW) followed by a neutral density filter, a linear polarizer and a quarter-wave plate producing circular
polarized light. All these components are summarized in figure~\ref{fig:1} as (a). The circular polarization 
of the laser light in combination with a second rotatable polarizer (b) allows to choose any angle of polarization without a change of intensity. 
After the polarizer the light beam passes the compensator (c). The latter two components are mounted on rotation tables. 
%
\begin{figure}[!!!h]
\begin{center}
	\includegraphics[width=\textwidth]{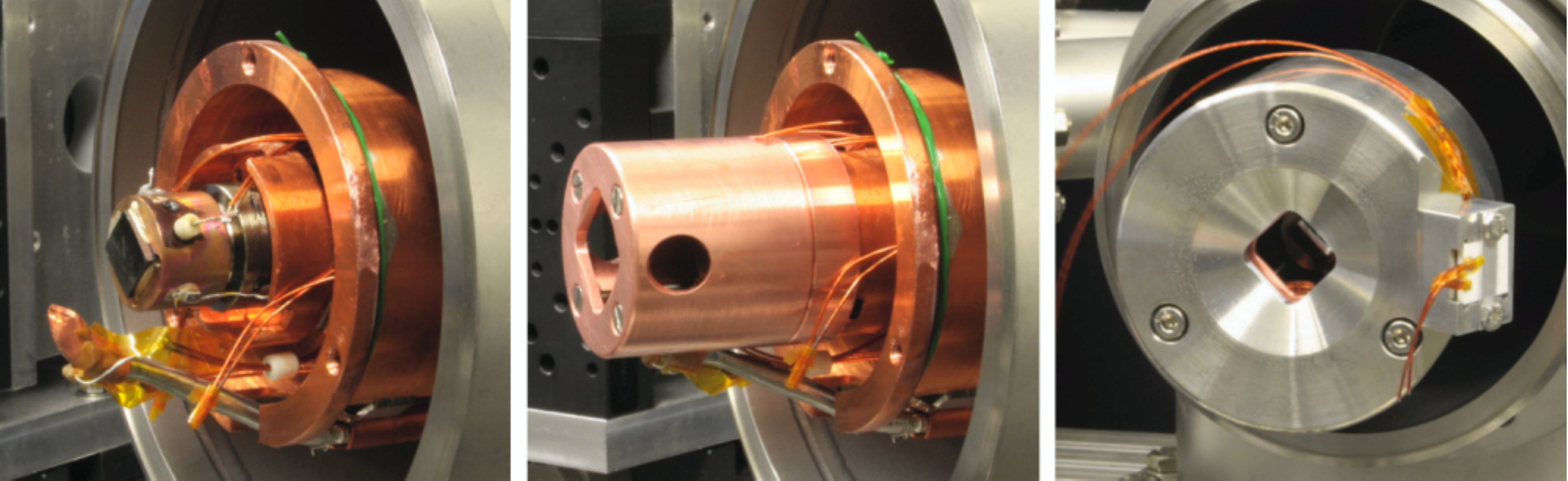} 
	\caption{Head of the cryostat containing the substrate with two cold shields and analyzer with detector.
	Left: HOPG substrate mounted on a copper block and the nozzle on the capillary for the gas inlet. Middle: Inner cold shield connected to the 1$^{\rm st}$ stage of the Gifford McMahon cold head with the openings for the ellipsometry laser. Right: Outer cold shield connected to the 2$^{\rm nd}$ stage
	of the Gifford McMahon cold head containing the analyzer and the detector. 
	}
	\label{fig:4}	
\end{center}
\end{figure}
The substrate (10~mm $\times$ 10~mm) is grade SPI-2 HOPG (SPI Supplies) with a mosaic angle of $0.8^{\circ} \pm 0.2^{\circ}$. The substrate is glued to a copper holder using electrically-conductive silver epoxy (Polytec H20E). The holder is cooled by a two-stage cryocooler of the Gifford McMahon-type (Sumitomo Heavy Industries Ltd., model RDK 408D). The first stage has a cooling power of 34~W at 40~K, the second stage provides a cooling power of 1~W at 4~K. The outer cold shield is connected to the first stage. The second stage cools the substrate and the inner cold shield. The temperature of the substrate, measured by a LakeShore DT-670B-SD temperature sensor mounted on the copper block that holds the substrate, can be set to an arbitrary value above 20~K by heating. To avoid birefringence of cold optical windows, both cold shields have free entrance and exit openings for the ellipsometry laser. At the exit opening of the outer cold shield the analyzer and the detector are mounted as shown in figure~\ref{fig:4}. The analyzer is a linear polarizer of 12~mm diameter and 0.28~mm thickness (Thorlabs LPVISB050, not laminated). The temperature at the analyzer is about 90~K (with a maximum gradient of 0.6~K/min during cool down.). The detector is a 9 $\times$ 9 mm$^2$  windowless Si-PIN photodiode (Hamamatsu S-3590-19) read out by a current amplifier (Femto DLPCA-200).
The other relevant optical components are the polarizer (PGT 2.05 Bernhard Halle Nachfolger GmbH - optische Werkst\"atten) and the compensator (CVI Melles Griot QWPM-543-04-4-R10)

After a full bake-out cycle of the set-up at a temperature of 423 K we prepared condensed films on the substrate. This was done by letting gas from a buffer volume with pressure of 2~mbar diffuse during short opening periods through a regulating valve (Pfeiffer UDV 146) and a heated capillary, which ends a few cm above the substrate surface. Different settings of the pressure in the buffer volume (typically 2~mbar) and the opening width and period of the dosing valve produce different step sizes in film thickness.
For our investigations we use standard krypton gas (purity 4.7). 

To clean the surface of the HOPG substrate a combination of resistive heating and laser ablation was used. The resistive heating was done by a TVO resistor up to 400~K. This temperature was held for about an hour before the laser ablation was started with a power density of 180~mW$\times$cm$^{-2}$ for 2-20~min. During the ablation the temperature was kept at 400~K.

The ablation set-up consists of a frequency doubled Nd:YAG laser (QUANTEL Brilliant), a Glan Laser polarizer (GL 10-A Thorlabs), high reflective mirrors and a beam homogenizer (SUSS CC-Q-300). The ablation laser provides pulses of 5~ns duration and of 200~mJ energy, at a repetition rate of 10 pulses per second. The power was reduced by a Glan laser polarizer to reach the desired power density at the substrate. The beam was homogenized by a high power beam homogenizer made of crossed cylindrical lens arrays to homogeneously illuminate the whole substrate.

To obtain the absolute start values the angle of the compensator was calibrated. This was done by using the ellipsometry laser, a linear polarizer, the compensator and another polarizer serving as analyzer. With the linear polarized light from the polarizer a circular polarized beam was prepared with the compensator. The residual linear polarization was measured by turning the analyzer by up to 180$^\circ$ and measure the intensity. The compensator was set to different angles to find the position, for which complete circular beam polarization could be achieved. The flattest measured curve defined the 45$^\circ$ position of the compensator corresponding to circular polarisation. 

Before each measurement series was started the gas chamber was baked at $\approx 423$ K to provide a clean environment. For each film the gas chamber was evacuated and filled with fresh gas to about 2~mbar. The purity of the used krypton was permanently monitored by a residual gas analyzer (RGA).

\begin{figure}[!!!h]
\centering
 \includegraphics[width=1.0\textwidth]{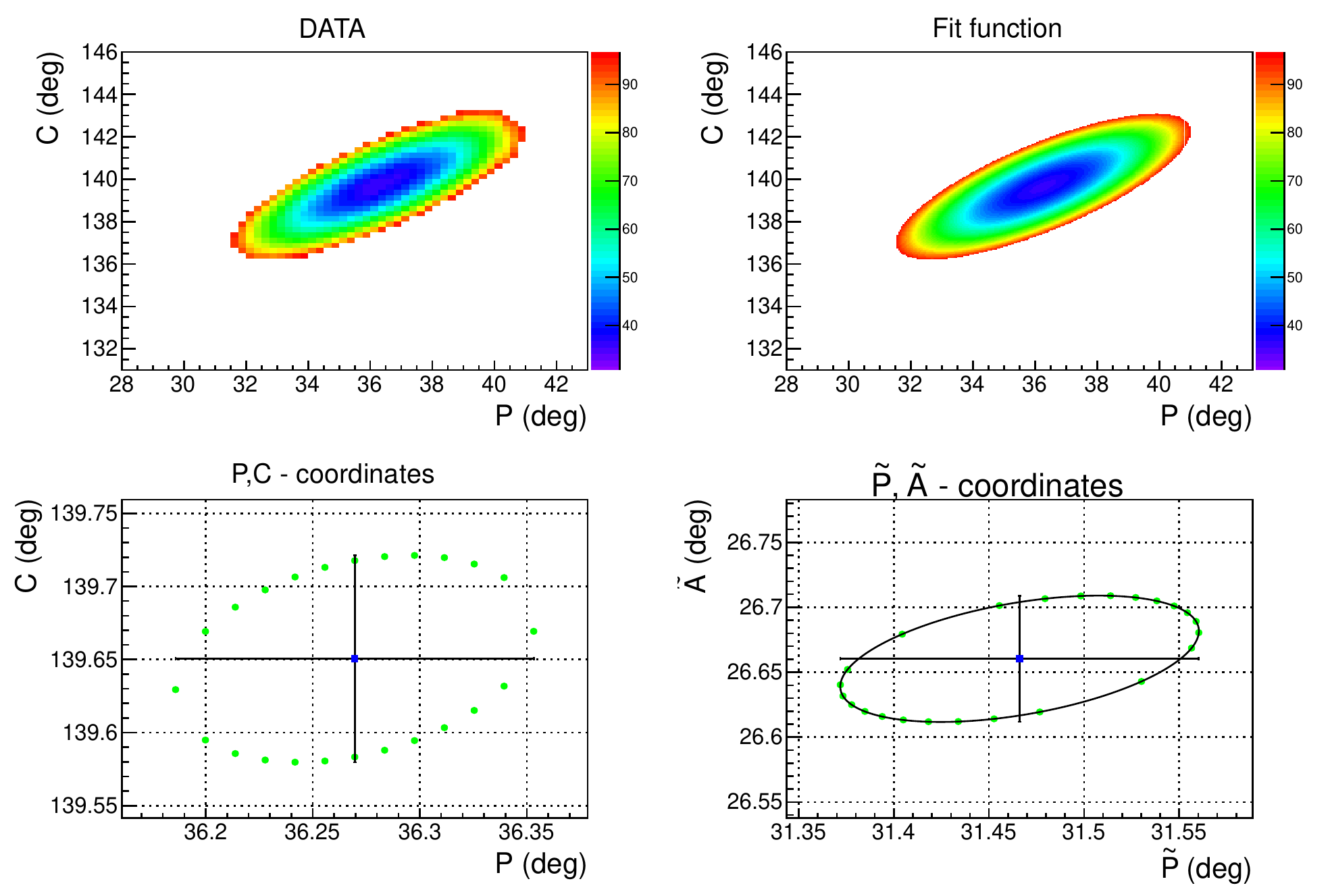} 
	\caption{Scan in the $PC$ plane of one condensation step and analysis at the example of measurement HAN3. The region around the intensity minimum of the data (upper left) was fitted with an elliptic paraboloid which is rotated in the $PC$ plane (upper right) yielding the minimum  $(P_\mathrm{m},C_\mathrm{m})$ with the corresponding errors (lower left). The green dots show the 1$\sigma$ error ellipse in the original (lower left) and in the transformed coordinates (lower right). The correlation and errors in the transformed coordinates were obtained from an elliptic fit to the transformed points (lower right).}
	\label{fig:Scan_fit}
\end{figure}

During the measurement it turned out that the iterative online minima search from the PA ellipsometry was not precise enough to find the correct minima because of the rather flat and broad PC minima (as shown in figure \ref{fig:pc_simulations}). Hence the area of $\pm$10$^{\circ}$ ($\pm$7.5$^{\circ}$) around the minima, found by the iterative method, was scanned with a typical step size of 1$^{\circ}$ (0.25$^{\circ}$). These data were analysed offline by fitting them, using a paraboloid locally around the minimum (see figure \ref{fig:Scan_fit}). 
The typical errors of the position in $PC$ coordinates are below  $0.1^{\circ}$. These errors are obtained from the least square fit and multiplied by $\sqrt{\chi^2_\mathrm{red}}$
\footnote{This correction using $\chi^2_\mathrm{red} = \chi^2/N_\mathrm{dof}$ 
($N_\mathrm{dof}$ is the number of degrees of freedom of the fit) is applied, since the size of the uncertainties of the intensity measurement is unknown but considered to be constant for all data points of a measurement.}. To obtain the uncertainties and the corresponding correlation after the transformation to $\tilde P \tilde A$ - coordinates, the points of the error ellipse in PC coordinates were also transformed and fitted (figure \ref{fig:Scan_fit} lower right).
This analysis was done for each condensed layer of krypton to obtain the intensity minima $(\tilde P_\mathrm{m},\tilde A_\mathrm{m})$, their uncertainties ($\Delta \tilde P_\mathrm{m}, \Delta \tilde A_\mathrm{m}$) and the correlation ($\tilde \rho$) for the whole condensation process.

\section{Experimental results}
\label{sec:results}

\begin{figure}[!!!htp]
	\centering
	\includegraphics[width=0.90\textwidth]{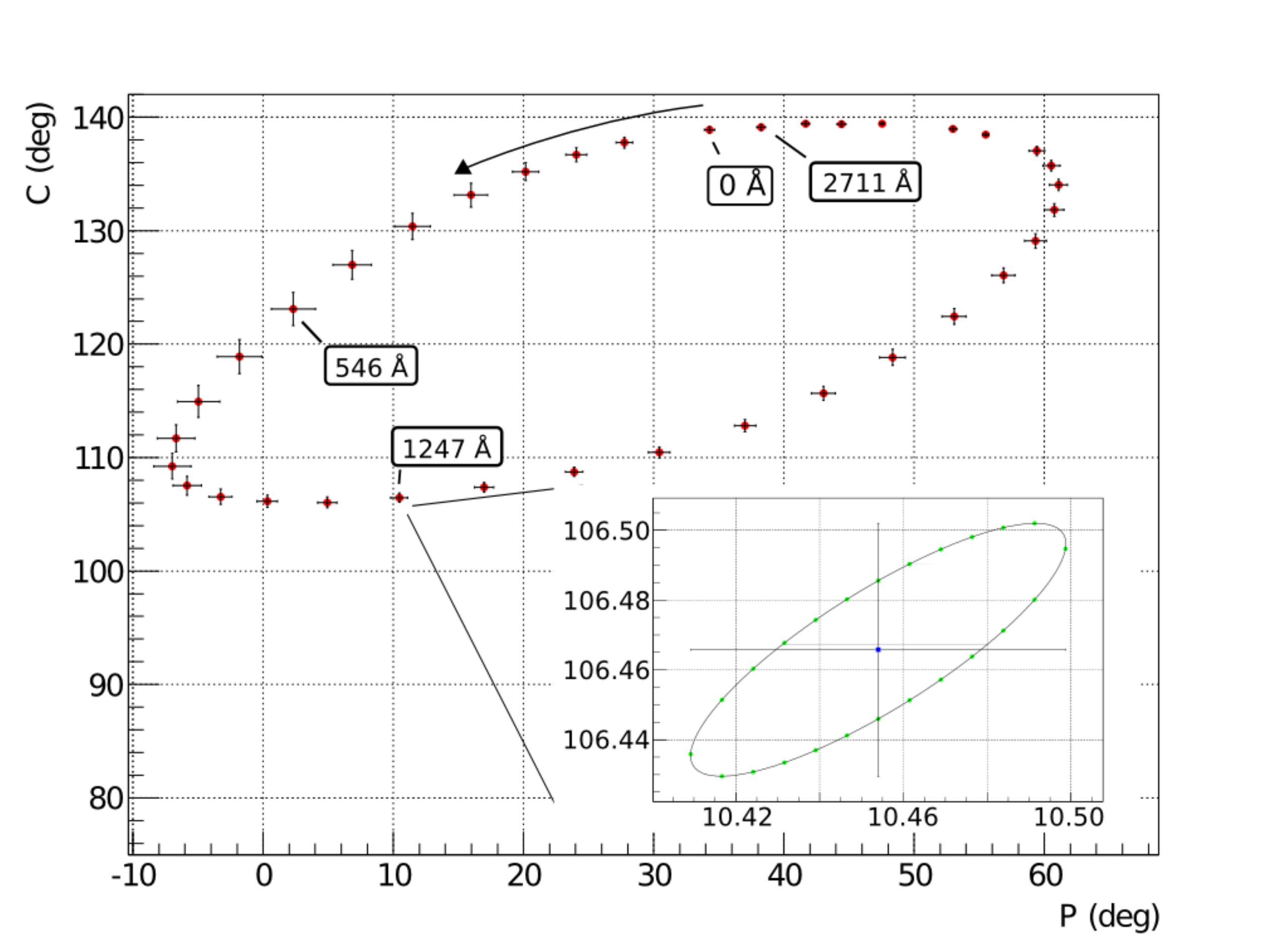}
	\caption{Intensity minima for polarizer and compensator $(P_\mathrm{m},C_\mathrm{m})$ of PC ellipsometry of measurement HA1. In this plot the results of the off-line fits are shown for one complete film. The errors are multiplied by a factor of 20 for the sake of clarity. The substrate was cleaned by a combination of heating and ablation. The inserted plot shows the data point of step 16 with the corresponding error ellipse. The size of the error bars changes over the whole film because the shape of the minima changes with increasing film thickness (see figure \ref{fig:pc_simulations}).}
	\label{fig:PC-ellipse}
\end{figure}

For the investigation of the PC ellipsometry three different measurement series (H, HA, HAN - for their meaning see further below) were carried out. For all measurements a krypton film of about 3000\AA\  thickness was condensed in about 15-30 steps. After each step a PC ellipsometry was carried out yielding an intensity minimum $(P_\mathrm{m},C_\mathrm{m})$. The results of a complete measurement is shown in figure \ref{fig:PC-ellipse}.
With the help of  equation~(\ref{eqn:pseudo_trans}) these minima were transformed into the minima $(\tilde P_\mathrm{m}, \tilde A_\mathrm{m})$ in $\tilde P \tilde A$ coordinates as shown in figure \ref{fig:Scan_fit}. 
All datasets were corrected  for a substrate tilted with respect to the plane of incident, as well as for angular offsets of the polarizer, compensator and analyzer. The uncertainties of these offsets were treated as systematic errors.

The conditions for the three measurement series are as follows (more details about differences between these measurement series will be given below in this section.):
\begin{enumerate}
\item \textbf{Measurements H1-H4:}\\ For these measurements the substrate was cleaned only by heating the substrate to about 400~K.
\item \textbf{Measurements HA1-HA3:}\\ A combination of heating of the substrate to about 400~K and ablation was used to clean the surface of the HOPG.
\item \textbf{Measurements HAN1-HAN6:}\\ To improve the reproducibility of the PC ellipsometry a new set of measurements was executed as for HA1-HA3, but with these differences:
\begin{enumerate}
\item The substrate was freshly cleaved 
\item P and C were scanned in a range of $\pm7.5^{\circ}$ with a step size of $0.25^{\circ}$ 
	instead of a range of $\pm 10^\circ$ with a step size of  $1^\circ$.
\item The heating temperature was raised to approx. 500~K before and during ablation
\end{enumerate}
\end{enumerate}

\begin{figure}[!!!htp]
	\centering
	\includegraphics[width=0.90\textwidth]{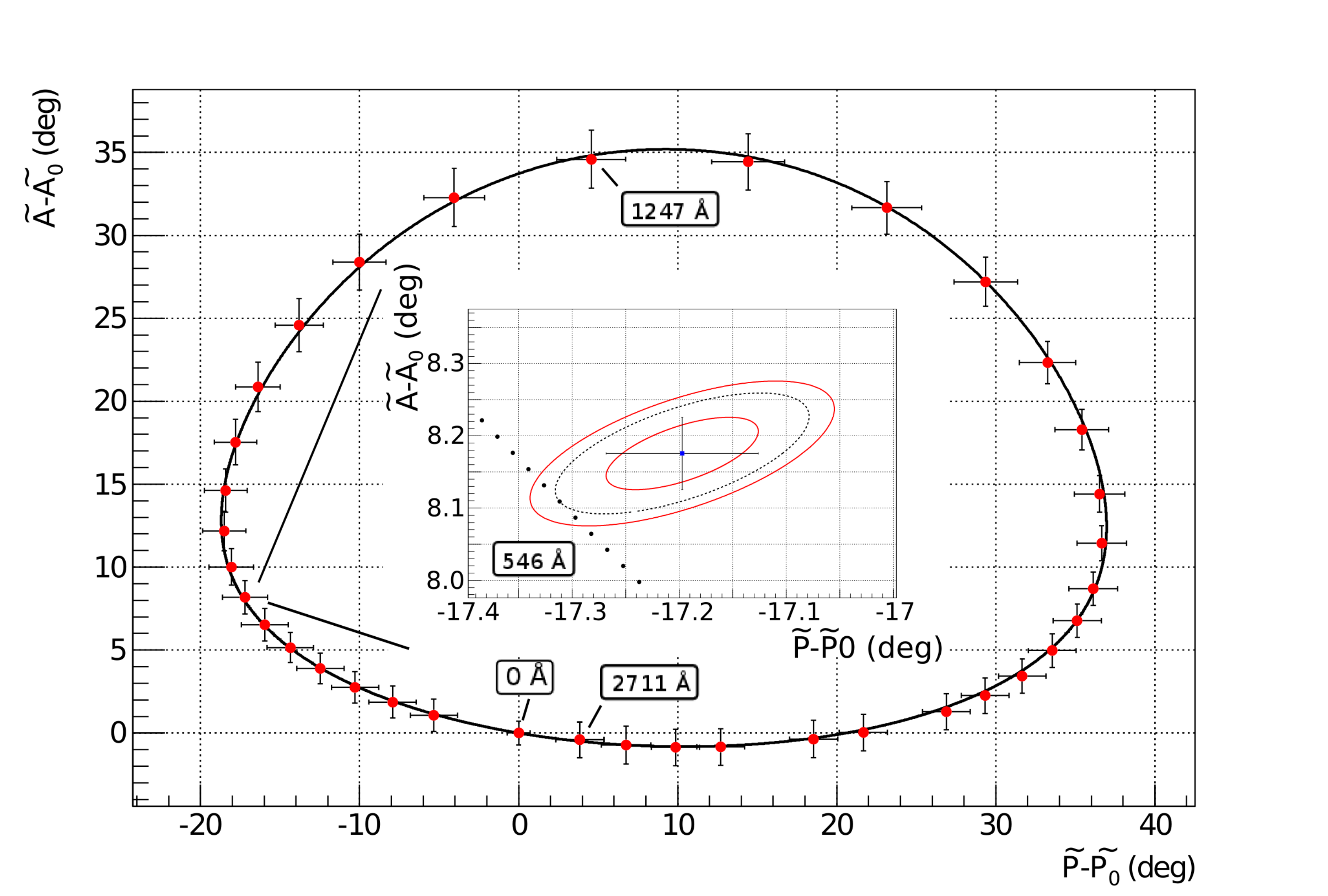}
	\caption{Intensity minima 
	$(\tilde P_\mathrm{m} - \tilde P_0, \tilde A_\mathrm{m} - \tilde A_0)$ of the  PC
	 ellipsometry of measurement HA1 transformed into $\tilde P \tilde A$ coordinates according to 
	 equation (\ref{eqn:pseudo_trans}). The line is a fit to the intensity minima were the parameters n$_1$, 
	 n$_2$ and k$_2$ were varied. The corresponding film thicknesses are marked by labels. 
	 The arrow denotes that the film thickness increases clockwise. 
	 The errors are multiplied by factor 20 for the sake of example. 
	 The inserted plot shows the data point of layer 7 together with the 1 sigma (red line), 
	 1.67 (black dashed) and 2 sigma error ellipse (red line)
	 to better illustrate how the distance $\chi_7^2$ is calculated with the help of equation 
	 (\ref{equ:Chi2}) (here $\chi^2_7 = 2.79 = 1.67^2$). 
	 The best fitting curve is shown by black dots were each dot has a distance of 1\AA.
	 It yields a $\chi^2_\mathrm{red} = 7.1$, which is accounted for by scaling the fit uncertainties (see text).}
	\label{fig:fig8}
\end{figure}

\medskip
The optical constants of HOPG $(n_2, k_2)$ as well as those of condensed krypton films $(n_1)$ were also obtained from a fit to the measured intensity minima $(\tilde P_\mathrm{m}, \tilde A_\mathrm{m})$ as shown in figure \ref{fig:fig8}.
First many theoretical curves $(\tilde P_\mathrm{m}, \tilde A_\mathrm{m})$ or $(P_\mathrm{m},  A_\mathrm{m})$ respectively, were calculated and compared to the measurements. This was done by varying the optical constants of HOPG $n_2, k_2$ and the index of refraction of krypton $n_1$. The angle of incident $\Phi_0$ and the absorption coefficient of krypton $k_1$ were kept fix during analysis. For each variation the ratio of the two reflection coefficients $\rho_S$  (equation (\ref{eqn:Grundgln})) was calculated in 1 \AA\ steps. From this the ($P,A$) values were derived using equation (\ref{eqn:pa_elli}). The curve with the smallest distance to the intensity minima 
$(\tilde P_\mathrm{m}, \tilde A_\mathrm{m})$ yields the values for the optical constants. The correct distance between an intensity minimum $(\tilde P_\mathrm{m}, \tilde A_\mathrm{m})$ and the fit curve normalized to the uncertainties is given by:

\begin{eqnarray}\label{equ:Chi2}
\chi_i^2 = &\frac{1}{(1-\tilde{\rho}^2_i)} \cdot \left [
      \frac{(\tilde P_{\mathrm{m},i} - P_{Fit,i})^2}{\sigma_{\tilde{P},i}^2} 
   + \frac{(\tilde A_{\mathrm{m},i} - A_{Fit,i})^2}{\sigma_{\tilde{A},i}^2}  \right.\\ \nonumber
& ~~~~~~~~~~~~~~~\left. - 2\cdot \tilde{\rho_i}\cdot \frac{(\tilde P_{\mathrm{m},i} - {P}_{Fit,i}) \cdot (\tilde A_{\mathrm{m},i} - {A}_{Fit,i})}{\sigma_{\tilde{P},i}\sigma_{\tilde{A},i}}\right ]~~~
\end{eqnarray}

The distance calculations takes also the correlation coefficient $\tilde \rho_i$ of the uncertainties 
$\Delta \tilde A_{\mathrm{m},i}$ and $\Delta \tilde P_{\mathrm{m},i}$ into account. The summed distance $\chi^2 = \sum_i \chi_i^2$ is minimized in the fit.
It should be noted that the fit uses relative values in $PA$ and $\tilde P \tilde A$ coordinates by subtracting
the corresponding values for zero film thickness $P_0, A_0$ and $\tilde P_0, \tilde A_0$ respectively.
We have not considered this detail in equation (\ref{equ:Chi2}) for the sake of convenience to read this equation.

\begin{figure}[!!!htp]
	\centering
	\includegraphics[width=0.90\textwidth]{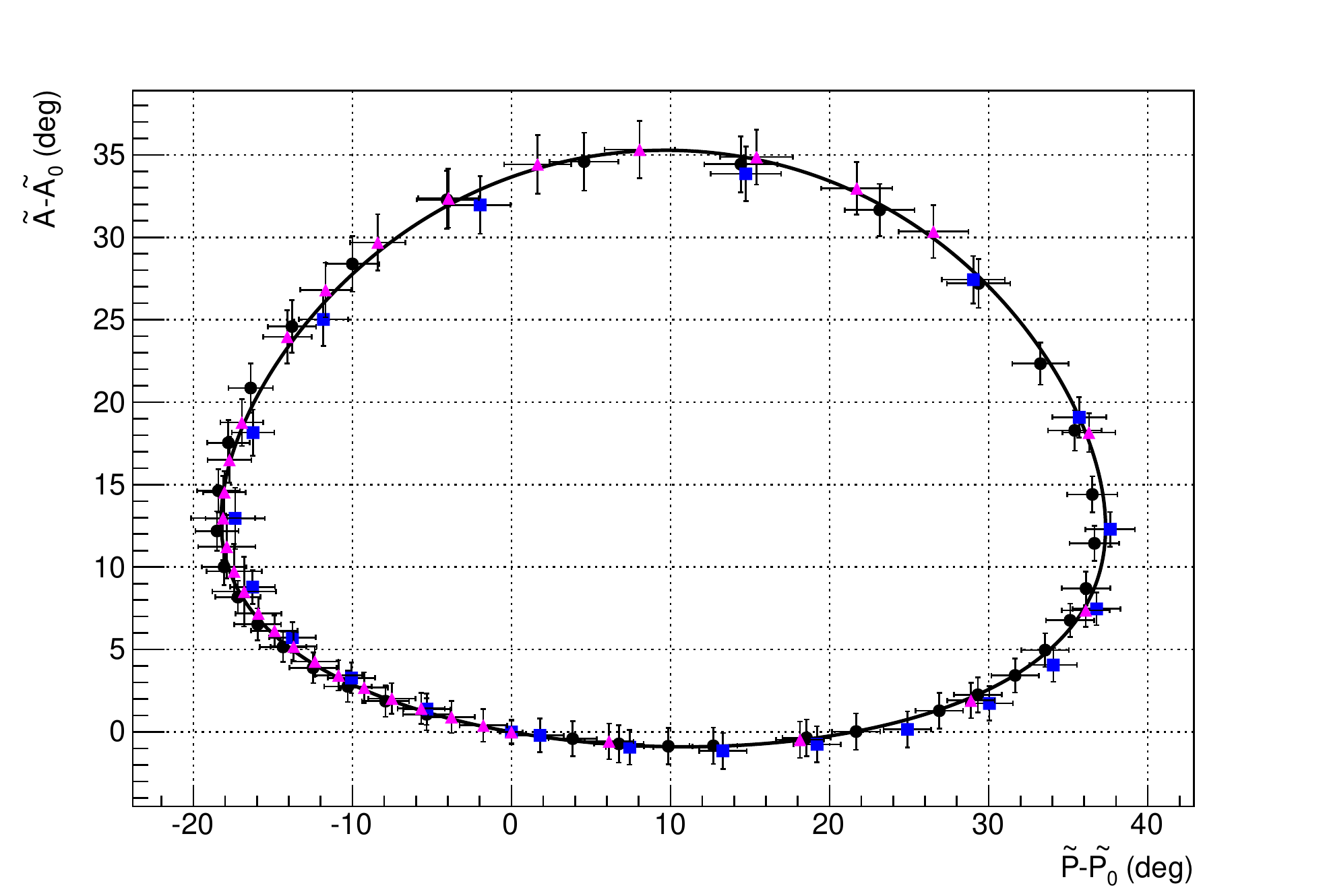}
	\caption{Results of three measurements HA1 - HA3. The solid line denotes the fit to all datasets and results in n$_1$=1.281, n$_2$=2.617 and k$_2$=1.001. The incident angle was fixed to 59.81$^{\circ}$ and k$_1$ to 0. All errors were multiplied by a factor of 20 for the sake of clarity. The fits yielding $\chi^2_\mathrm{red} (\mathrm{HA1}) = 7.1, ~\chi^2_\mathrm{red} (\mathrm{HA2}) = 13.6, ~\chi^2_\mathrm{red} (\mathrm{HA3}) = 4.9$.}
	\label{fig:all_meas}
\end{figure}

The fit to the corresponding angles $(\tilde A_\mathrm{m} - \tilde A_0, \tilde P_\mathrm{m} - \tilde P_0)$ yields the refractive indices of the condensed krypton film $n_1$ and the optical constants of the
HOPG substrate $n_2, k_2$ (see table~\ref{tab:tab2}). These fits like shown in figure \ref{fig:fig8} yield $3 < \chi^2_\mathrm{red} < 22$ which point to unrecognised systematic errors. Possible sources of these are a possible surface roughness, impurities, porosity of the film. Unfortunately our not complete knowledge of these 
effects does not allow a correction of these systematics. To account for their influence on the fit results, we  scale the fit uncertainties $\Delta n_1, \Delta n_2, \Delta k_2$ with $sqrt{\chi^2_\mathrm{red}}$. Although HOPG is an anisotropic material we describe it by only one isotropic refraction index instead of ordinary and extraordinary refractive indices. The results of all three measurements of series HA are shown in figure \ref{fig:all_meas}. 
 
\medskip
In addition to the HA-series some HAN-series measurements were performed, in order to see whether our method could be improved further by increasing the heating temperature of the substrate to about 500 K. The increased substrate temperature during ablation can help to get a cleaner surface because the heat conductivity of HOPG out of plane decreases and the conductivity in plane increases with increasing temperature \cite{Issi}. Therefore, the energy of the ablation laser pulses are transferred less into the substrate and is dissipated mostly at the surface. In addition, the step size of the $P$- and $C$-scanning was reduced to $0.25^\circ$ to obtain
a more precise minimum.

\begin{figure}[!!!htp]
\centering
	\includegraphics[width=0.90\textwidth]{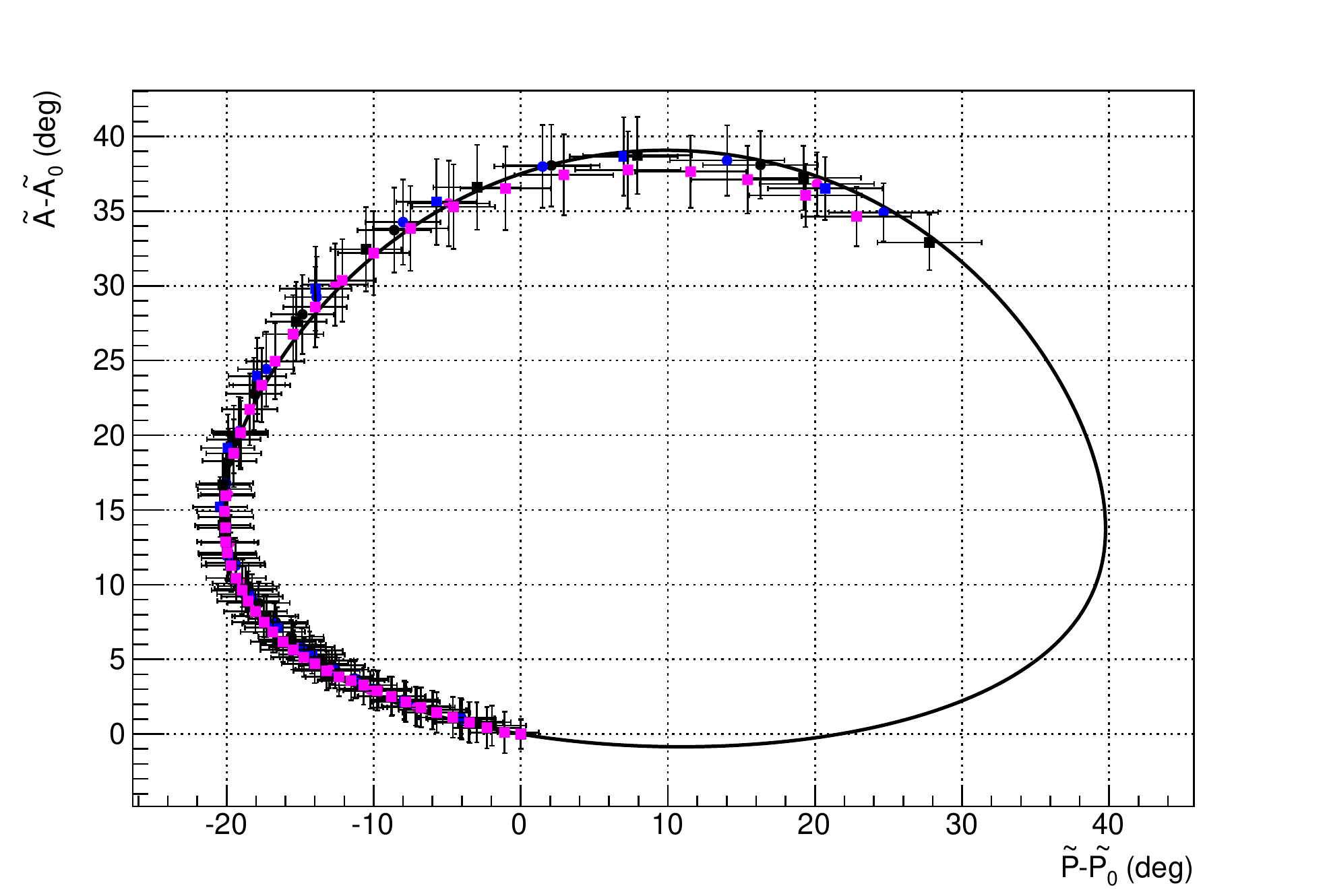}
	\caption{Results of six measurements HAN1 - HAN6. The solid line denotes the fit to all datasets and results in n$_1$=1.272, n$_2$=2.698 and k$_2$=0.813. The incident angle was fixed to 61.19$^{\circ}$ and k$_1$ to 0. All errors were multiplied by a factor of 20 for the sake of clarity. The last parts (thicknesses higher than approx. 1400\AA) of all datasets were not analysed because of the influence of a too high residual gas pressure of water in the set-up. The fits yielding $\chi^2_\mathrm{red}$ between 3.2 and 21.5 .}
	\label{fig:all_HAN}
\end{figure}

Both changes lead to a higher reproducibility of the measured data but resulted in a much longer time span to measure one complete condensation of a film up to 3000~\AA~thickness. Unfortunately the
residual gas pressure of water in the vacuum chamber was higher during this measurement series 
thus the data could not be analysed for very large film thickness, because water was condensing onto the film. The whole condensing procedure took more than 24 hours due to the elaborate minimum search 
caused by the fine step size. This condensing water caused a non-closed curve in $PC$ or $\tilde P \tilde A$ coordinates for a krypton film. Therefore we limited our analysis to film thickness up to 1400~\AA~only.
The results of all analysed data points are shown in figure \ref{fig:all_HAN}.
 
 \clearpage

\begin{figure}[!!!h]
\centering
	\includegraphics[width=0.8\textwidth]{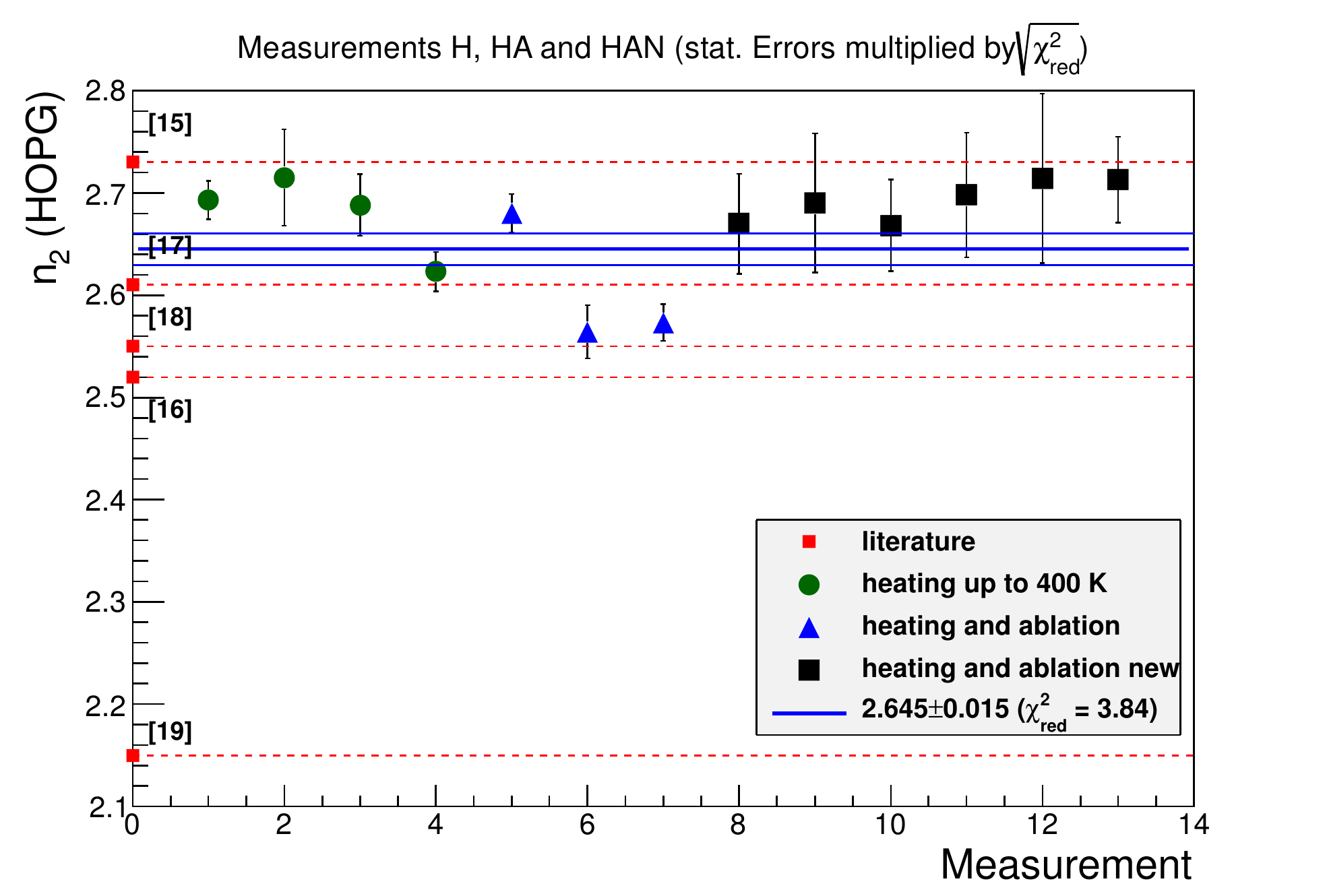} 
	\includegraphics[width=0.8\textwidth]{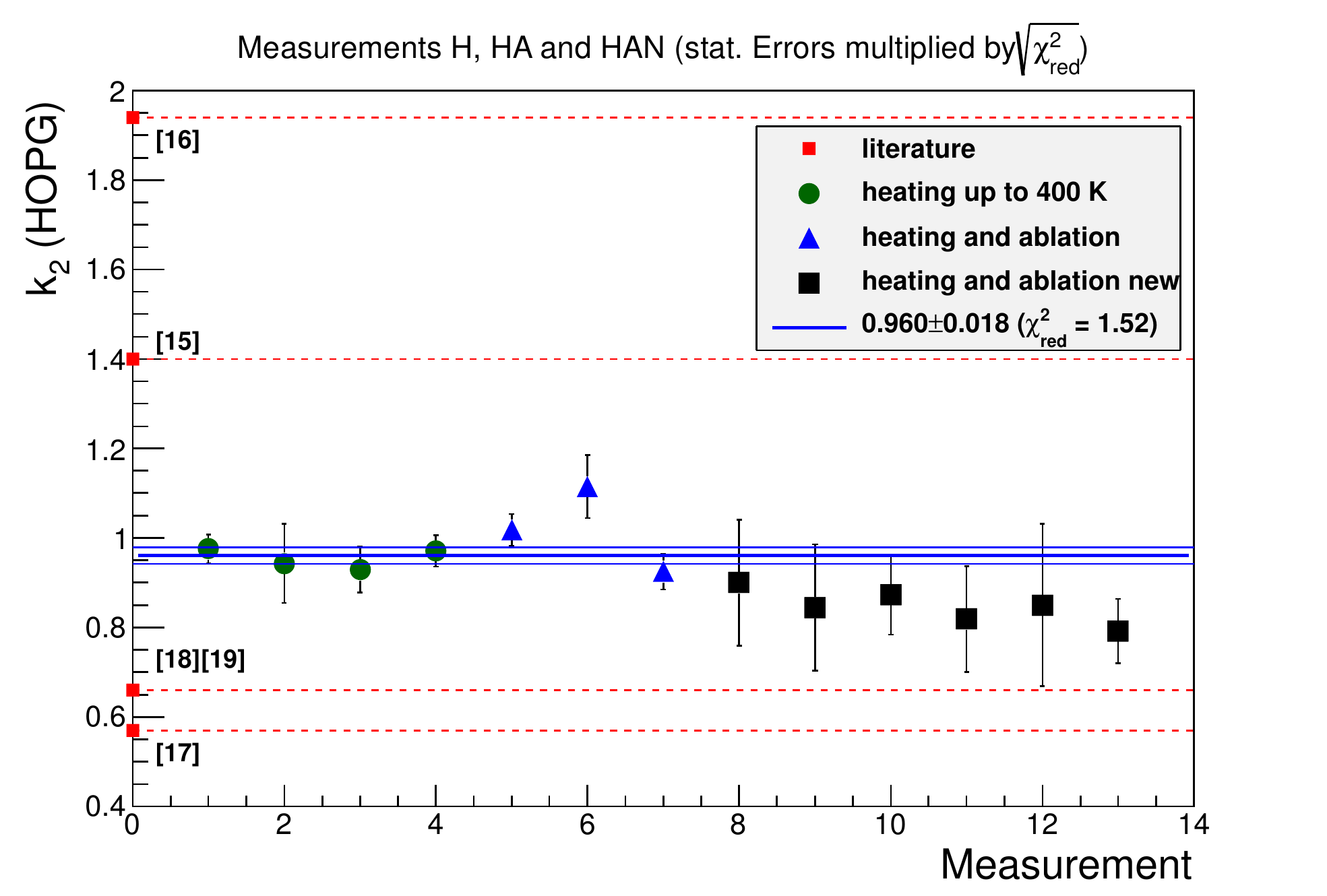} 
	\caption{The index of refraction $n_2$ (top panel) and the extinction coefficient $k_2$ (bottom panel) of HOPG were determined by PC ellipsometry after different methods of cleaning the substrate were applied. The dashed lines denote the values found in literature \cite{Gre69, jellison2007, Taft, Berman, Ergun}. The solid blue lines show the mean value obtained from a fit to all measurements and the corresponding one sigma errors. The uncertainties of this average were obtained by scaling the fit error with $\sqrt{\chi^2_\mathrm{red}}$}
	\label{fig:N2_HOPG}
\end{figure}

\begin{figure}[!!!h]
\centering
	\includegraphics[width=0.8\textwidth]{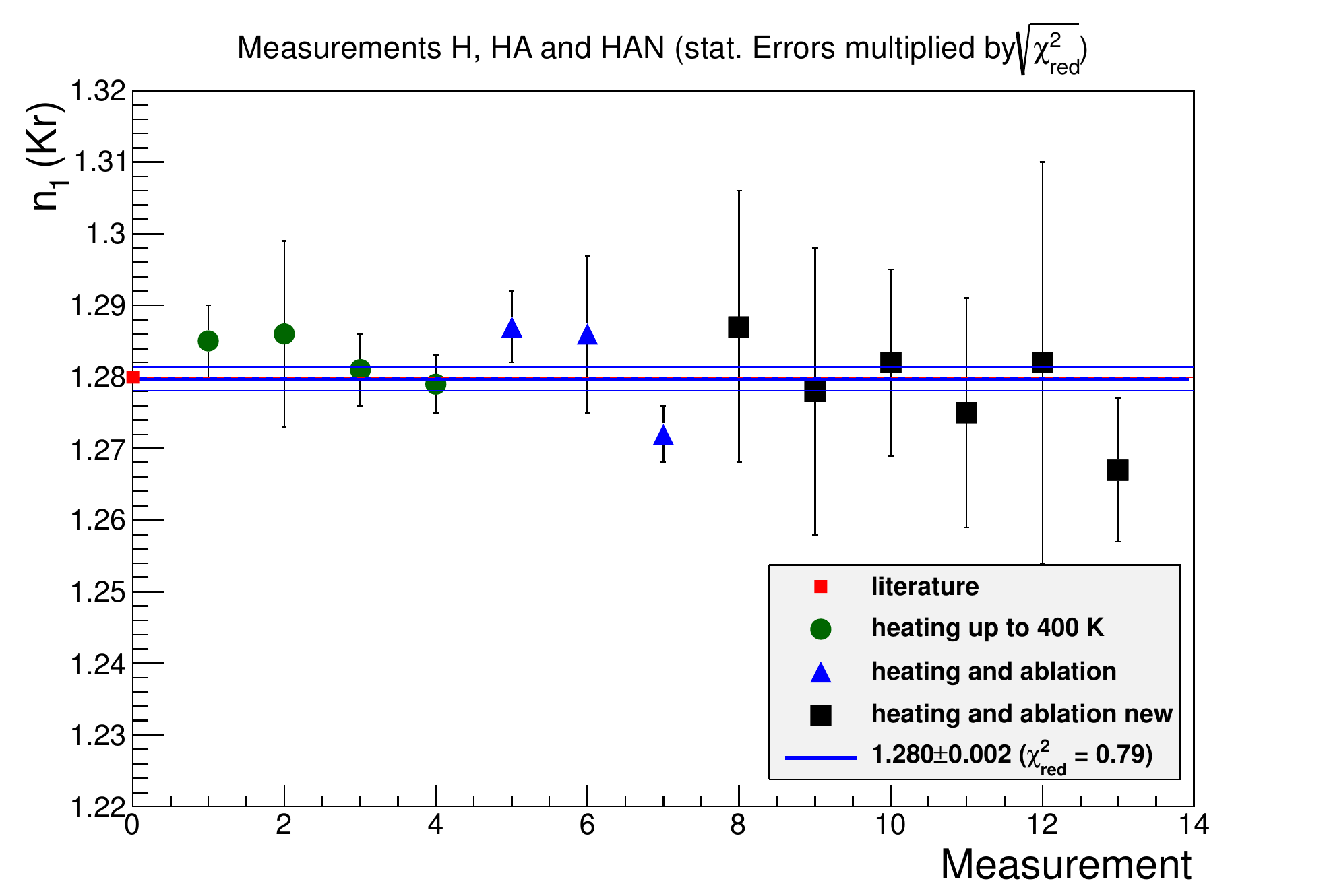} 
	\caption{The index of refraction $n_1$ of the condensed krypton film determined by PC ellipsometry after different methods of cleaning the substrate were applied. The dashed line denote the values of Kruger \cite{Kruger}. The solid blue lines shows the mean value obtained from a fit to all measurements and the corresponding one sigma errors (see text). }
	\label{fig:N1_Kr}
\end{figure}

\medskip
Our previous investigations showed that the combination of laser ablation and heating leads to the most stable measurements in comparison to other cleaning methods \footnote{We would like to note, that the proper cleaning not only affects the index of refraction of HOPG but also those of the krypton film.}.
Figures \ref{fig:N2_HOPG} and \ref{fig:N1_Kr} show the results of the optical parameters $n_2,k_2$
of HOPG and $n_1$ of krypton from all of our measurements. The results from the measurements within a series and between the three measurements series are consistent with each other but still the average exhibit higher $\chi^2_\mathrm{red}$ than expected. To account for this we scale  the fit error with $\sqrt{\chi^2_\mathrm{red}}$. These results 
from figures \ref{fig:N2_HOPG} and \ref{fig:N1_Kr} 
are compared with literature values   in table~\ref{tab:tab2}.

\medskip

The refractive index $N_1$ of solid krypton deviates from most of the literature values. In the Mainz Neutrino Mass Experiment a similar deviation of the refractive index of tritium was observed, measured by ellipsometry as well as by measurements of the energy loss of electrons in quench condensed deuterium films \cite{aseev2000}. This observation was explained by a porous film due to the quench condensation of the tritium gas. So we calculate the density of the kryton films with the help of Clausius Mosotti`s equation:
\begin{equation}
\frac{{N_1}^2 - 1}{{N_1}^2 + 2} = \frac{{n_1}^2 - 1}{{n_1}^2 + 2} = \frac{4 \pi}{3} \alpha_{pol} \frac{N_A}{V_{mol}} 
\label{eqn:mosotti}
\end{equation}
and the measured index of refraction the density of our krypton film:
\begin{equation}
\rho_{Kr}= \frac{{n_1}^2 - 1}{{n_1}^2 + 2} \cdot \frac{3 M}{4\pi \alpha_{pol} N_A} ~.
\label{eqn:rho}
\end{equation}

Using the polarizability $\alpha_{pol} = 2.46 \cdot 10^{-24}$~cm$^3$ \cite{Kit05} and the molar mass ($M$) of krypton we obtain a density of our krypton film of:
\[
\rho_{Kr}= 2.369\frac{g}{cm^3}.
\]

This is about 23\% less than the expected value of $\rho= 3.081 \frac{g}{cm^3}$ \cite{Beaumont}. Therefore the film can be assumed to be porous due to the quench condensation of the krypton gas, as in the
case of the quench condensed tritium and deuterium films \cite{aseev2000}.

The results show a reasonable agreement of our measured optical refractive indices for krypton and HOPG with values reported in the literature, for both krypton and HOPG (see column 3 of Table \ref{tab:tab2}). Specially one finds: $N_2 = 2.73 - i \cdot 1.4$~(at 633~nm)\cite{Gre69},  $N_2 = 2.52 - i \cdot 1.94$ ~(at 546~nm) \cite{jellison2007},  $N_2 =  2.61- i \cdot 0.57$~(633~nm)\cite{Taft} taken from \cite{Ergun}, $N_2 = 2.55 - i \cdot 0.66$~(at 633~nm) \cite{Berman} and $N_2 = 2.15 - i \cdot 0.66$~(at 541~nm)\cite{Ergun} yielding the ranges $n_2 = 2.15 - 2.73$ and $k_2 = 0.57 - 1.94$ (see table \ref{tab:tab2}). For the refractive index of krypton $k_1$ only the results of Kruger \cite{Kruger} are shown because most of the other values in literature are measured with different techniques and thus lead to different results. Due to the quench condensation of krypton the film might be porous and this might lead to a lower index of refraction. The values obtained from different methods e.g. measurements with refractometers leads to a index of refraction of n$_1$=1.375 \cite{Sin79}.

\begin{table}[!!!h]
\centering
\begin{tabular}{p{2.0cm}cc}
\hline 
refractive	& this work & literature\\
index & \\
\hline \hline
$n_1$ & $1.282 \pm 0.008$ & 1.28*\\
$n_2$ & $2.645 \pm 0.030$ &  2.15--2.73\\
$k_2$ &$0.964 \pm 0.037$ & 0.57--1.94\\
\hline
\end{tabular}
\caption{Refractive indices of krypton $N_1 = n_1$ and the HOPG substrate $N_2 = n_2 - i k_2$. The second column gives the results from PC-ellipsometry measurements at $T=23$~K (the error is the quadratic sum of statistic and systematic error). The third column presents a range of literature values. *For the refractive index of krypton $k_1$ only the results of Kruger \cite{Kruger} are shown because it is also measured with quench condensed films.
 }
\label{tab:tab2}
\end{table}	


%
%

\section{Conclusion and outlook}
\label{ausblick}
%
In this paper, we presented a new variant of ellipsometry, which we dubbed PC-ellipsometry since
the polarizer and the compensator are rotated to find the intensity minimum, where the analyzer behind the substrate remains at fixed angle. This method allows a simple polarisation analysis close to the substrate  
even at cryogenic temperatures inside a vacuum chamber. We determined
the optical constants of condensed krypton and of the HOPG substrate that are consistent with literature values.
We demonstrated that thicknesses of condensed krypton films up to 3000 \AA~can be determined.
Our results show that PC ellipsometry can reach accuracies similar to that of standard PA ellipsometry. 
We propose a transformation of the intensity minima $(P_\mathrm{m},C_\mathrm{m})$ of PC ellipsometry into $\tilde P \tilde A$  coordinates, which allows one use the same evaluation tools as for PA ellipsometry.\\
%
The described ellipsometry set-up is designed to operate at the KATRIN experiment. Due to the temperature of 77~K at the site of operation inside a superconducting split-coil magnet, the use of vacuum windows with direct sight onto the substrate is impossible. We will use the new method to carry out the polarisation analysis inside the setup. The incoming polarized laser beam will be guided by one or two mirrors onto the substrate. The mirrors will be placed behind the polarizer and the compensator onto the substrate. It should be noted that in general dielectric mirror coating do influence the polarisation state of any incident light beam. However, for our high-precision mirrors (Laseroptik 11028J1) we found that these particular coatings preserved a defined polarization state well enough to perform accurate PC ellipsometry.

\section*{Acknowledgment}

The KATRIN experiment is supported by the Bundesministerium f\"ur Bildung und Forschung (BMBF) under the contract number 05A08PM1.  

\newpage
\section*{Appendix A}

\begin{table}[htbp]
\begin{center}
\begin{tabular}{ccccccc} 
Meas. & $\alpha (^{\circ})$ & $n_1$ & $n_2$ & $k_2$ & $\chi^2_\mathrm{red}$\\ 
\hline 
H1 & $59.81$ & ($1.285\pm0.005$)  & ($2.693\pm0.019$) & ($-0.976\pm0.032$) & 7.1\\ 
H2 & $59.81$ & ($1.286\pm0.013$)  & ($2.715\pm0.047$) & ($-0.943\pm0.088$) & 45.5\\ 
H3 & $59.81$ & ($1.281\pm0.005$)  & ($2.688\pm0.030$) & ($-0.929\pm0.051$) & 25.5\\ 
H4 & $59.81$ & ($1.279\pm0.004$)  & ($2.623\pm0.019$) & ($-0.971\pm0.035$) & 4.6\\ 
\\
HA1 & $59.81$ & ($1.287\pm0.005$) & ($2.680\pm0.019$) & ($-1.018\pm0.035$) & 7.1 \\ 
HA2 & $59.81$ & ($1.286\pm0.011$) & ($2.564\pm0.026$) & ($-1.115\pm0.070$) & 13.6 \\ 
HA3 & $59.81$ & ($1.272\pm0.004$) & ($2.573\pm0.018$) & ($-0.925\pm0.040$) & 4.9 \\ 
\\
HAN1 & $61.19$ & ($1.287\pm0.019$)  & ($2.670\pm0.049$) & ($-0.900\pm0.141$) & 14.5\\ 
HAN2 & $61.19$ & ($1.278\pm0.020$)  & ($2.690\pm0.068$) & ($-0.844\pm0.141$) & 16.2\\ 
HAN3 & $61.19$ & ($1.282\pm0.013$)  & ($2.668\pm0.045$) & ($-0.873\pm0.089$) & 3.2\\ 
HAN4 & $61.19$ & ($1.275\pm0.016$)  & ($2.698\pm0.061$) & ($-0.819\pm0.118$) & 16.5\\ 
HAN5 & $61.19$ & ($1.282\pm0.028$)  & ($2.714\pm0.083$) & ($-0.850\pm0.181$) & 21.5\\ 
HAN6 & $61.19$ & ($1.267\pm0.010$)  & ($2.713\pm0.042$) & ($-0.792\pm0.072$) & 10.6\\ 

\end{tabular} 
\end{center}
\caption{All fitted values for the different measurement series. The incident angle was measured and fixed for the fit. Krypton was assumed to be non absorbent thus the absorption coefficient was $k_1$=0 in the analysis. The errors are statistical only and are multiplied by $\sqrt{\mathcal{X}^2}$. }
\end{table}

\begin{table}[htbp]
\begin{center}
\begin{tabular}{ccccccccccc}
 d~(\AA) & P~($^{\circ}$) & C~($^{\circ}$) & $\Delta P~(^{\circ})$ & $\Delta C~(^{\circ})$  & $\rho$ & $\tilde{P}~(^{\circ})$ & $\tilde{A}~(^{\circ})$ & $\Delta\tilde{P}~(^{\circ})$ & $\Delta\tilde{A}~(^{\circ})$  & $\tilde{\rho}$ \\ \hline
0 & 34.570 & 138.672 & 0.069 & 0.051 & 0.560 & 30.799 & 27.275 & 0.059 & 0.036 & 0.185 \\ 
122 & 28.020 & 137.499 & 0.068 & 0.051 & 0.557 & 25.467 & 28.346 & 0.058 & 0.033 & 0.180 \\ 
188 & 24.288 & 136.389 & 0.068 & 0.051 & 0.562 & 22.882 & 29.145 & 0.058 & 0.031 & 0.189 \\ 
255 & 20.455 & 134.940 & 0.071 & 0.054 & 0.586 & 20.515 & 30.034 & 0.059 & 0.031 & 0.205 \\ 
326 & 16.146 & 132.774 & 0.070 & 0.053 & 0.577 & 18.331 & 31.166 & 0.058 & 0.029 & 0.286 \\ 
397 & 11.634 & 129.992 & 0.068 & 0.051 & 0.560 & 16.440 & 32.430 & 0.057 & 0.028 & 0.413 \\ 
470 & 7.028 & 126.644 & 0.074 & 0.057 & 0.619 & 14.847 & 33.805 & 0.057 & 0.032 & 0.480 \\ 
546 & 2.363 & 122.604 & 0.070 & 0.053 & 0.585 & 13.602 & 35.451 & 0.054 & 0.035 & 0.615 \\ 
623 & -1.740 & 118.451 & 0.072 & 0.055 & 0.606 & 12.734 & 37.287 & 0.052 & 0.041 & 0.642 \\ 
704 & -4.908 & 114.472 & 0.072 & 0.055 & 0.607 & 12.290 & 39.463 & 0.050 & 0.048 & 0.631 \\ 
782 & -6.569 & 111.298 & 0.072 & 0.056 & 0.606 & 12.377 & 41.903 & 0.049 & 0.054 & 0.574 \\ 
860 & -6.886 & 108.785 & 0.068 & 0.051 & 0.562 & 13.000 & 44.813 & 0.048 & 0.059 & 0.540 \\ 
935 & -5.702 & 107.128 & 0.069 & 0.052 & 0.566 & 14.407 & 48.135 & 0.053 & 0.065 & 0.445 \\ 
1011 & -3.014 & 106.270 & 0.070 & 0.052 & 0.570 & 17.001 & 51.865 & 0.060 & 0.071 & 0.358 \\ 
1083 & 0.555 & 105.904 & 0.068 & 0.051 & 0.560 & 20.799 & 55.660 & 0.069 & 0.076 & 0.311 \\ 
1162 & 5.216 & 105.830 & 0.069 & 0.051 & 0.559 & 26.749 & 59.551 & 0.083 & 0.080 & 0.281 \\ 
1247 & 11.035 & 106.397 & 0.069 & 0.051 & 0.561 & 35.364 & 61.864 & 0.098 & 0.080 & 0.258 \\ 
1334 & 17.392 & 107.266 & 0.070 & 0.052 & 0.564 & 45.222 & 61.704 & 0.105 & 0.077 & 0.209 \\ 
1421 & 23.987 & 108.490 & 0.071 & 0.053 & 0.570 & 53.949 & 58.936 & 0.099 & 0.070 & 0.107 \\ 
1506 & 30.724 & 110.302 & 0.073 & 0.054 & 0.587 & 60.164 & 54.476 & 0.088 & 0.064 & -0.008 \\ 
1592 & 37.382 & 112.767 & 0.068 & 0.051 & 0.556 & 64.049 & 49.612 & 0.075 & 0.053 & -0.077 \\ 
1672 & 43.113 & 115.445 & 0.072 & 0.054 & 0.583 & 66.210 & 45.563 & 0.070 & 0.050 & -0.114 \\ 
1763 & 48.883 & 118.937 & 0.068 & 0.051 & 0.557 & 67.313 & 41.682 & 0.064 & 0.041 & -0.125 \\ 
1846 & 53.390 & 122.378 & 0.070 & 0.053 & 0.575 & 67.447 & 38.714 & 0.063 & 0.039 & -0.066 \\ 
1936 & 57.336 & 126.117 & 0.068 & 0.051 & 0.561 & 66.917 & 35.980 & 0.060 & 0.035 & 0.016 \\ 
2007 & 59.618 & 129.014 & 0.069 & 0.052 & 0.569 & 65.917 & 34.048 & 0.060 & 0.035 & 0.097 \\ 
2083 & 61.021 & 131.763 & 0.069 & 0.052 & 0.571 & 64.347 & 32.242 & 0.059 & 0.036 & 0.159 \\ 
2154 & 61.467 & 134.014 & 0.069 & 0.051 & 0.565 & 62.462 & 30.706 & 0.058 & 0.037 & 0.203 \\ 
2224 & 60.744 & 135.629 & 0.070 & 0.053 & 0.572 & 60.118 & 29.531 & 0.059 & 0.039 & 0.215 \\ 
2289 & 59.535 & 136.863 & 0.070 & 0.053 & 0.573 & 57.696 & 28.563 & 0.059 & 0.040 & 0.221 \\ 
2411 & 55.732 & 138.312 & 0.070 & 0.052 & 0.565 & 52.467 & 27.302 & 0.059 & 0.042 & 0.216 \\ 
2477 & 53.021 & 138.713 & 0.071 & 0.053 & 0.573 & 49.344 & 26.919 & 0.060 & 0.043 & 0.209 \\ 
2594 & 47.782 & 139.274 & 0.069 & 0.052 & 0.566 & 43.492 & 26.435 & 0.059 & 0.042 & 0.213 \\ 
2650 & 45.052 & 139.345 & 0.071 & 0.052 & 0.572 & 40.658 & 26.413 & 0.060 & 0.041 & 0.191 \\ 
2711 & 41.899 & 139.264 & 0.076 & 0.056 & 0.605 & 37.556 & 26.552 & 0.063 & 0.043 & 0.147 \\ 
\end{tabular}
\end{center}
\caption{Calculated film thickness and the corresponding ($P, C, \Delta P, \Delta C, \rho$) and ($\tilde{P},\tilde{A}, \Delta \tilde{P}, \Delta \tilde{A}, \tilde{\rho}$) values for measurement HA1.}
\label{Tabelle_HA1}
\end{table}

\end{document}